\documentclass[12pt, draftclsnofoot, onecolumn]{IEEEtran}

\ifCLASSINFOpdf
  \usepackage[pdftex]{graphicx}
  \DeclareGraphicsExtensions{.pdf,.jpeg,.png}
\else
  \usepackage[dvips]{graphicx}
\fi
\usepackage{epstopdf}
\usepackage{amssymb,amsmath,amsthm}
\usepackage{amssymb}
\usepackage{wasysym}
\usepackage{algorithm}
\usepackage{algorithmic}
\usepackage{array}
\usepackage{cite}
\usepackage{color}
\usepackage{url}

\usepackage{epsfig,latexsym}
\usepackage{flushend}
\usepackage{cite}
\usepackage{verbatim}
\usepackage{amsopn}
\usepackage{booktabs}

\usepackage{stfloats}
\usepackage{amsmath}

\usepackage{subfigure}
\usepackage{hyperref}
\usepackage{mathrsfs}
\usepackage{setspace}

\begin{document}

\title{Compressed Sensing Based Channel Estimation for Movable Antenna Communications}
\author{{Wenyan Ma, \IEEEmembership{Student Member, IEEE}, Lipeng Zhu, \IEEEmembership{Member, IEEE}, and  Rui Zhang, \IEEEmembership{Fellow, IEEE}}
	\vspace{-25pt}
	
	\thanks{W. Ma and L. Zhu are with the Department of Electrical and Computer Engineering, National University of Singapore,
		Singapore 117583 (Email: {wenyan@u.nus.edu}, {zhulp@nus.edu.sg}).
		
		R. Zhang is with School of Science and Engineering, Shenzhen Research Institute of Big Data, The Chinese University of Hong Kong, Shenzhen, Guangdong 518172, China (e-mail: rzhang@cuhk.edu.cn). He is also with the Department of Electrical and Computer Engineering, National University of Singapore, Singapore 117583 (e-mail: elezhang@nus.edu.sg).
}}
\maketitle

\begin{abstract}
In this letter, we study the channel estimation for wireless communications with movable antenna (MA), which requires to reconstruct the channel response at any location in a given  region where the transmitter/receiver is located based on the channel measurements taken at finite locations therein, so as to find the MA's location for optimizing the communication performance. To reduce the pilot overhead
and computational complexity for channel estimation, we propose a new successive transmitter-receiver compressed sensing (STRCS) method by exploiting the efficient representation of the channel responses in the given transmitter/receiver region (field) in terms of multi-path components. Specifically, the field-response information (FRI) in the angular domain,
including the angles of departure (AoDs)/angles of arrival (AoAs) and complex coefficients of all significant multi-path components are sequentially estimated based on a finite number of channel measurements taken at random/selected locations by the MA at the transmitter and/or receiver. Simulation results demonstrate that the proposed channel reconstruction method outperforms the benchmark schemes in terms of both pilot  overhead and channel reconstruction accuracy.

\end{abstract}
\begin{IEEEkeywords}
	Movable antenna (MA), channel estimation, field-response information (FRI), compressed sensing.
\end{IEEEkeywords}

\section{Introduction}
With the development of multiple-input multiple-output (MIMO) technologies, the capacity of wireless communication systems has experienced a significant boost, owing to the exploitation of new degrees of freedom (DoFs) in the spatial domain \cite{lu2014an,zhu2019milli}. However, conventional MIMO and/or massive MIMO systems employ fixed-position antennas (FPAs), which may suffer from deep  fading at a given time and/or frequency resource block (RB). To address this issue, movable antenna (MA) was recently proposed as a promising technology for efficiently adapting to the spatial channel variation  \cite{zhu2022modeling,ma2022mimo,zhu2023movable}. With the aid of mechanical drivers, MAs can be maneuvered flexibly within a continuous spatial region where the transmitter/receiver is located to improve the wireless channel condition. 

Prior studies have demonstrated the effectiveness of MAs in enhancing communication performance \cite{zhu2022modeling,ma2022mimo,zhu2023movable}. In \cite{zhu2022modeling}, the MA architecture and field-response-based channel model were proposed for the single-MA system, where the signal-to-noise ratio (SNR) gain over its FPA counterpart was investigated under both deterministic and stochastic channels. In \cite{ma2022mimo}, an MA-based MIMO communication system was analyzed, where the channel capacity was maximized via joint optimization of transmitter and receiver MAs' positions. It was shown that the MA-based MIMO systems can achieve a substantial gain in channel capacity compared to conventional FPA-based MIMO systems. Besides, the MA-based multiuser communication system was considered in \cite{zhu2023movable}, where the MAs' positions of multiple users were jointly optimized to minimize the total transmit power, subject to the minimum achievable rate requirement of each user. It was revealed that the antenna position optimization can significantly decrease the total transmit power as compared to conventional FPA systems with perfect/imperfect channel state information (CSI). 

It is worth noting that in practice, the moving speed of MAs may be constrained by the mechanical drivers. As such, MAs  are most suitable for communication scenarios with slow channel variations over time (e.g., when the transmitter/receiver is deployed at fixed location) and limited bandwidth (i.e., narrow-band transmission), such as machine-type communications (MTC) \cite{shariatmadari2015machine}. In such scenarios, conventional diversity techniques in time and/or frequency domain may not be applicable, and thus MA can be a new and cost-effective solution to enhance the spatial diversity, especially when the number of antennas/radio-frequency (RF) chains at the transceiver is small due to cost consideration.

For MA communication systems,  the knowledge of CSI in the entire region for positioning the MAs is essential for finding their best locations to optimize the communication performance. This thus gives rise to a new channel estimation problem for reconstructing the complete CSI in a given region (field) based on finite channel measurements taken by the MA at random/selected locations therein. However, conventional channel estimation methods for MIMO systems with FPAs only need to acquire CSI at given  positions where the antennas are located (see, e.g., \cite{lee2016channel,wang2020compressed}), and thus cannot be directly applied to solve the new channel estimation problem for MA systems. On the other hand, if the complete CSI is obtained by positioning the MA over all possible locations in the given region (e.g., by spatially sampling the region with a sufficiently high rate) to perform channel measurement, this will result in prohibitively high pilot overhead and large data transmission  delay. 

To efficiently solve the channel estimation problem  for MA systems, we propose in this letter a new method by exploiting the sparse representation of the channel responses in the MA's residing field in terms of  multi-path components (MPCs), thus named field-response information (FRI). Specifically, the FRI in the angular domain,
including the angles of departure (AoDs)/angles of arrival (AoAs) and complex coefficients of all significant MPCs at the transmitter/receiver are first recovered based on finite channel measurements taken by the MA at random/selected locations. Then, the channel response between any two locations in the MA's residing regions at the transmitter and receiver, respectively, is reconstructed based on the estimated angular-domain FRI. Furthermore, to reduce the overhead of FRI estimation, we propose a new successive  transmitter-receiver  compressed sensing (STRCS) scheme which is implemented in three steps. First, the transmitter-side MA (T-MA) moves over different locations  and the receiver-side MA (R-MA) at a fixed position  measures the corresponding  channels, based on which the estimation of the transmitter-side MPCs (T-MPCs)' AoDs is formulated as a sparse signal recovery problem and solved by compressed sensing algorithm. Second, with the T-MA's position fixed,  the R-MA moves over different locations while measuring the corresponding channels for estimating  the AoAs of the receiver-side MPCs (R-MPCs), similarly as that for the AoD estimation. Finally, based on the obtained AoD and AoA information, the T-MA and R-MA are jointly moved for estimating the complex coefficients of all MPCs by using the least square (LS)-based  algorithm\footnote{Note that the proposed method also applies to the case where FPA is deployed at the transmitter or receiver, while only the MA needs to be moved over different locations when conducting the channel estimation.}. Simulation results show that the proposed channel reconstruction method is more efficient than benchmark schemes in terms of pilot  overhead and yet achieves higher channel reconstruction accuracy.

\textit{Notations}: Symbols for vectors (lower case) and matrices (upper case) are in boldface.  $(\cdot)^T $, $(\cdot)^* $, and $(\cdot)^H $ denote the transpose, conjugate, and conjugate transpose (Hermitian), respectively. The sets of $P\times{Q}$ dimensional complex and real matrices are denoted by $\mathbb{C}^{P\times{Q}}$ and $\mathbb{R}^{P\times{Q}}$, respectively. We use $\boldsymbol{a}[p]$ and $\boldsymbol{A}[p,q]$ to denote the $p$th entry of vector $\boldsymbol{a}$ and the entry of matrix $\boldsymbol{A}$ at the $p$th row and $q$th column, respectively. $\otimes$ denotes Kronecker product. We use $\textrm{diag}(\boldsymbol{a})$ to denote the square diagonal matrix with the elements of vector $\boldsymbol{a}$ on the main diagonal. $\|\boldsymbol{a}\|_2$ and $\|\boldsymbol{A}\|_F$ denote the $l_2$ norm of vector $\boldsymbol{a}$ and Frobenius norm of matrix $\boldsymbol{A}$, respectively. The vectorization of matrix $\boldsymbol{A}$ is denoted by $\textrm{vec}(\boldsymbol{A})$. $\boldsymbol{A}^\dagger$ denotes pseudo inverse.

\section{System and Channel Model}
\subsection{System Model}

\begin{figure}[!t]
	\centering
	\includegraphics[width=130mm]{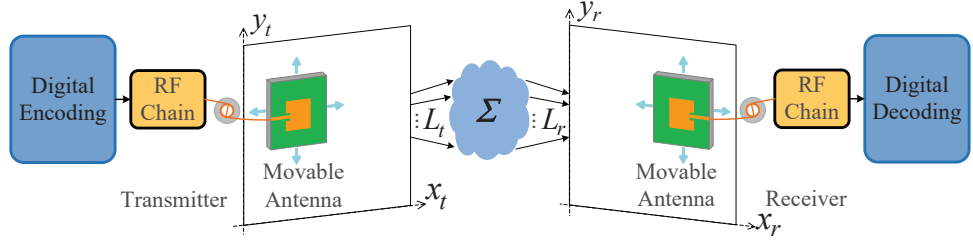}
	\caption{Wireless communication with one single MA at both the transmitter and receiver.}
	\label{FIG1}
\end{figure}

As shown in Fig.~\ref{FIG1}, we consider an MA communication system with one single MA at the transmitter as well as at the receiver. The positions of T-MA and R-MA can be flexibly adjusted, with their coordinates denoted by $\boldsymbol{t}=[x^t, y^t]^T \in \mathcal{C}^t$ and $\boldsymbol{r}=[x^r, y^r]^T \in \mathcal{C}^r$, respectively. For simplicity, we assume that $\mathcal{C}^t$ and $\mathcal{C}^r$ are given 2D regions at the transmitter/receiver side and for convenience, we further assume that they  are both square regions of equal size $A \times A$. 

Let $h(\boldsymbol{t},\boldsymbol{r})$ denote the channel from the transmitter to the receiver, which depends on MAs' positions $\boldsymbol{t}$ and $\boldsymbol{r}$. Let $s$ represent the transmit signal with normalized power equal to one. The received signal can be expressed as
\begin{equation}\label{y}
  y(\boldsymbol{t}, \boldsymbol{r}) = \sqrt{P} h(\boldsymbol{t}, \boldsymbol{r}) s + z,
\end{equation}
where $P$ is the transmit power and $z \sim \mathcal{CN}(0,\sigma^2)$ denotes the additive white Gaussian noise (AWGN) at the receiver with $\sigma^2$ being the average noise power. For notation simplicity and without loss of generality, we set $s=1$ as the pilot signal for FRI estimation.

\subsection{Field-Response Based Channel Model}

For the considered MA communication system, the channel between the transmitter and receiver is determined by the propagation environment and $\{\boldsymbol{t}, \boldsymbol{r}\}$. To characterize the FRI in the MA's residing region, we consider the far-field wireless channel model, which assumes that the size of $\mathcal{C}^t$/$\mathcal{C}^r$ is much smaller than the distance between the transmitter and receiver  \cite{zhu2022modeling,ma2022mimo}. Thus, the AoD, AoA and amplitude of the complex coefficient of each MPC can be assumed to be approximately constant in $\mathcal{C}^t$ and $\mathcal{C}^r$, while the phase of each MPC's complex coefficient generally varies over  different T-/R-MA's positions in $\mathcal{C}^t$/$\mathcal{C}^r$.

The number of T-MPCs and R-MPCs are denoted by $L^t$ and $L^r$, respectively\footnote{Note that in conventional multi-path channel modeling, the number of multi-paths usually refers to that at the receiver side, i.e., $L^r$.}. For the $p$th ($p=1,2,\ldots,L^t$) T-MPC, its elevation and azimuth AoDs are represented by $\tilde{\theta}^t_p \in [0, \pi]$ and $\tilde{\phi}^t_p \in [0, \pi]$, respectively. For simplicity, the virtual AoDs are defined as $\theta^t_p \triangleq \sin \tilde{\theta}^t_p \cos \tilde{\phi}^t_p$ and $\phi^t_p \triangleq \cos \tilde{\theta}^t_p$. Then, the phase difference corresponding to the $p$th T-MPC between MA location $\boldsymbol{t}\in\mathcal{C}^t$ and the center of $\mathcal{C}^t$ is $2\pi\rho^t_p(\boldsymbol{t})/\lambda$, where $ \rho^t_p(\boldsymbol{t}) = x^t \theta^t_p + y^t \phi^t_p$ and $\lambda$ is the carrier wavelength. The transmitter-side field response vector (T-FRV) is thus obtained as \cite{zhu2022modeling,ma2022mimo,zhu2023movable}
\begin{equation}\label{g}
  \boldsymbol{g}(\boldsymbol{t}) = \left[ e^{j\frac{2\pi}{\lambda}\rho^t_1(\boldsymbol{t})}, e^{j\frac{2\pi}{\lambda}\rho^t_2(\boldsymbol{t})}, \ldots, e^{j\frac{2\pi}{\lambda}\rho^t_{L^t}(\boldsymbol{t})} \right]^T \in{\mathbb{C}^{L^t}}.
\end{equation}

Similarly, let $\tilde{\theta}^r_q \in [0, \pi]$ and $\tilde{\phi}^r_q \in [0, \pi]$ denote the elevation and azimuth AoAs of the $q$th ($q=1,2,\ldots,L^r$) R-MPC, respectively. We also define the virtual AoAs as $\theta^r_q \triangleq \sin \tilde{\theta}^r_q \cos \tilde{\phi}^r_q$ and $\phi^r_q \triangleq \cos \tilde{\theta}^r_q$ for convenience. Then, the receiver-side field response vector (R-FRV) is obtained as \cite{zhu2022modeling,ma2022mimo,zhu2023movable}
\begin{equation}\label{f}
  \boldsymbol{f}(\boldsymbol{r}) \triangleq \left[ e^{j\frac{2\pi}{\lambda}\rho^r_1(\boldsymbol{r})}, e^{j\frac{2\pi}{\lambda}\rho^r_2(\boldsymbol{r})}, \ldots, e^{j\frac{2\pi}{\lambda}\rho^r_{L^r}(\boldsymbol{r})} \right]^T \in{\mathbb{C}^{L^r}},
\end{equation}
where $\rho_r^q(\boldsymbol{r}) = x^r \sin \theta^r_q + y^r \phi^r_q$ represents the difference in signal propagation distance for the $q$th R-MPC between position $\boldsymbol{r}\in\mathcal{C}^r$ and the center of $\mathcal{C}^r$.

Furthermore, let $\boldsymbol{\Sigma} \in{\mathbb{C}^{L^r \times L^t}}$ denote the path response matrix (PRM) characterizing  all T-/R-MPCs' responses between the centers of $\mathcal{C}^t$ and $\mathcal{C}^r$, where $\boldsymbol{\Sigma}[q,p]$ is the channel response coefficient between the $p$th T-MPC and the $q$th R-MPC. Thus, the end-to-end channel between the transmitter and the receiver as a function of the T-/R-MA's position $\boldsymbol{t}$/$\boldsymbol{r}$ can be expressed as
\begin{equation}\label{H}
  h(\boldsymbol{t}, \boldsymbol{r}) = \boldsymbol{f}(\boldsymbol{r})^H \boldsymbol{\Sigma} \boldsymbol{g}(\boldsymbol{t}).
\end{equation}
For MA systems, it is infeasible to estimate $h(\boldsymbol{t}, \boldsymbol{r})$ directly for all $\boldsymbol{t}\in\mathcal{C}^t$ and $\boldsymbol{r}\in\mathcal{C}^r$ by moving a pair of MAs over the whole regions of $\mathcal{C}^t$ and $\mathcal{C}^r$, respectively. However, it is observed from \eqref{H} that the channel $h(\boldsymbol{t}, \boldsymbol{r})$ between any $\{\boldsymbol{t}, \boldsymbol{r}\}$ can be reconstructed based on the angular-domain FRI, including  $\{\theta^t_p, \phi^t_p\}_{p=1}^{L^t}$, $\{\theta^r_q, \phi^r_q\}_{q=1}^{L^r}$, and $\boldsymbol{\Sigma}$, which are constant over $\mathcal{C}^t$ and $\mathcal{C}^r$. Hence, we propose to estimate the FRI components first, and then reconstruct $h(\boldsymbol{t}, \boldsymbol{r})$ over the entire $\mathcal{C}^t$ and $\mathcal{C}^r$ based on the estimated FRI via \eqref{H}. However, since we consider single MA at both the transmitter and receiver, only one pilot can be transmitted/received at each time for measuring the channel between a given pair of T-/R-MAs' positions. Moreover, the joint estimation of all FRI components is complicated and may incur prohibitively high computational complexity. Thus, efficient FRI estimation methods need to be designed to reduce the number of pilots (training time) as well as the computational complexity.

\section{Proposed Method for FRI  Estimation}
\begin{figure}
	\centering
	\includegraphics[width=110mm]{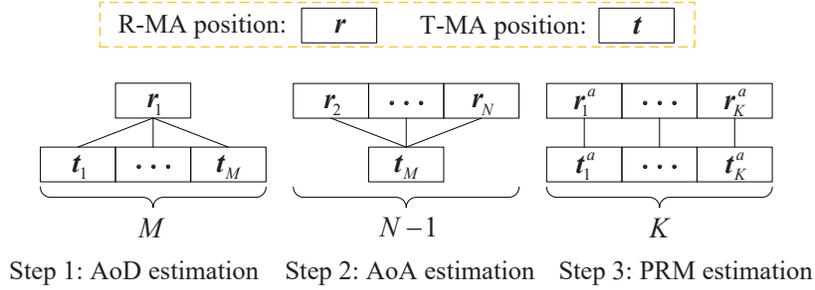}
	\caption{Illustration of the proposed STRCS method for angular-domain FRI estimation with MAs.}
	\label{FIG2}
\end{figure}

In this section, we present the new STRCS method to efficiently estimate the angular-domain FRI with affordable complexity, which is implemented in three steps as shown in Fig.~\ref{FIG2}. First, the T-MA moves over different locations  and the R-MA at a fixed position performs channel measurement for estimating the T-MPCs' AoDs, which is formulated as a sparse signal recovery problem and solved by compressed sensing algorithm. Second, the T-MA's position is fixed and the R-MA moves over different locations while measuring the corresponding channels to estimate the R-MPCs' AoAs, similarly as that for the AoD estimation. Finally, the T-MA and R-MA are jointly moved based on the obtained AoD and AoA information to estimate the complex coefficients of all MPCs using the LS-based algorithm.

\subsection{AoD Estimation}
To estimate the T-MPCs' AoDs efficiently, the single T-MA needs to be moved over different positions to send pilot signals. The receiver measures the corresponding channels based on the received pilot signals and then estimates the T-MPCs' AoDs. Let $M$ denote the total number of moved positions for the T-MA, which are denoted by $\{\boldsymbol{t}_m\}_{m=1}^{M}$ with $\boldsymbol{t}_m \triangleq [x^t_m, y^t_m]^T \in\mathcal{C}^t$. For each $\boldsymbol{t}_m$, we assume that one pilot is sent by the T-MA. Since the R-MA's position is fixed as $\boldsymbol{r}_1 \in\mathcal{C}^r$, in this step, the received signals are given by
\begin{align}
	y_m^t = \sqrt{P}\boldsymbol{f}(\boldsymbol{r}_1)^H \boldsymbol{\Sigma} \boldsymbol{g}(\boldsymbol{t}_m) + z_m^t, ~~m=1,2,\ldots,M,
\end{align}
where $z_m^t$ is the AWGN at the receiver and $z_m^t \sim \mathcal{CN}(0,\sigma^2)$. For convenience, we stack the $M$ received signals to obtain
\begin{align}
	\left(\boldsymbol{y}^t\right)^T = \sqrt{P}\boldsymbol{f}(\boldsymbol{r}_1)^H \boldsymbol{\Sigma} \boldsymbol{G}(\tilde{\boldsymbol{t}})  + \left(\boldsymbol{z}^t\right)^T,
\end{align}
with 
\begin{align}\label{yzGS}
	\boldsymbol{y}^t &\triangleq [y_1^t, y_2^t, \ldots, y_M^t]^T \in \mathbb{C}^{M}, \\
	\boldsymbol{z}^t &\triangleq [z_1^t, z_2^t, \ldots, z_M^t]^T \in \mathbb{C}^{M}, \notag \\
	\tilde{\boldsymbol{t}} &\triangleq \left[\boldsymbol{t}_1, \boldsymbol{t}_2, \ldots, \boldsymbol{t}_M\right] \in \mathbb{R}^{2 \times M}, \notag \\
	\boldsymbol{G}(\tilde{\boldsymbol{t}}) &\triangleq \left[\boldsymbol{g}\left(\boldsymbol{t}_1\right), \boldsymbol{g}\left(\boldsymbol{t}_2\right), \ldots, \boldsymbol{g}\left(\boldsymbol{t}_M\right)\right] \in \mathbb{C}^{L^t \times M}. \notag
\end{align}

As can be observed from \eqref{yzGS}, each column of $\boldsymbol{G}(\tilde{\boldsymbol{t}})$ is the T-FRV with a fixed AoD and different positions of T-MA, while each row of it can be regarded as the steering vector with a fixed position of T-MA and varying AoDs. Thus, for any AoD pair $\{\theta^t, \phi^t\}$, the transmitter-side steering vector can be expressed as
\begin{align}
	\boldsymbol{a}(\theta^t, \phi^t) \triangleq \left[ e^{-j\frac{2\pi}{\lambda}\tau^t_1(\theta^t, \phi^t)}, e^{-j\frac{2\pi}{\lambda}\tau^t_2(\theta^t, \phi^t)}, \ldots, e^{-j\frac{2\pi}{\lambda}\tau^t_{M}(\theta^t, \phi^t)} \right]^T \in{\mathbb{C}^{M}},
\end{align}
where $\tau^t_m(\theta^t, \phi^t) = x^t_m \theta^t + y^t_m \phi^t$ represents the signal propagation distance difference with respect to $\{\theta^t, \phi^t\}$. Then, the transmitter-side steering matrix can be obtained as
\begin{align}
	\boldsymbol{A}(\boldsymbol{\alpha}^t)\triangleq\left[\boldsymbol{a}(\theta^t_1, \phi^t_1), \boldsymbol{a}(\theta^t_2, \phi^t_2), \ldots, \boldsymbol{a}(\theta^t_{L^t}, \phi^t_{L^t})\right] \in \mathbb{C}^{M \times L^t},
\end{align}
with $\boldsymbol{\alpha}^t \triangleq [\theta^t_1, \theta^t_2, \ldots, \theta^t_{L^t}, \phi^t_1, \phi^t_2, \ldots, \phi^t_{L^t}]^T \in{\mathbb{R}^{2L^t}}$. According to the definition $\boldsymbol{A}(\boldsymbol{\alpha}^t) = \boldsymbol{G}(\tilde{\boldsymbol{t}})^H$, the received signal vector can be rewritten as
\begin{align}\label{yt2}
	\boldsymbol{y}^t = \sqrt{P} \boldsymbol{A}(\boldsymbol{\alpha}^t) \boldsymbol{\Sigma}^H \boldsymbol{f}(\boldsymbol{r}_1)   + \boldsymbol{z}^t \triangleq \boldsymbol{A}(\boldsymbol{\alpha}^t) \boldsymbol{x}^t   + \boldsymbol{z}^t,
\end{align}
with $\boldsymbol{x}^t \triangleq \sqrt{P}\boldsymbol{\Sigma}^H \boldsymbol{f}(\boldsymbol{r}_1) \in{\mathbb{C}^{L^t}}$. To efficiently estimate $\{\theta^t_p, \phi^t_p\}_{p=1}^{L^t}$ by compressed sensing, we make an approximation by uniformly discretizing $\theta\in[-1,1]$ and $\phi\in[-1,1]$ into $G$ grids with $G\gg L^t$ \cite{wang2020compressed}. Thus, we have $\boldsymbol{A}(\boldsymbol{\alpha}^t) \boldsymbol{x}^t \approx \bar{\boldsymbol{A}} \bar{\boldsymbol{x}}^t$, where $\bar{\boldsymbol{A}} \in{\mathbb{C}^{M\times{G^2}}}$
is an over-complete matrix and its $(g_1+(g_2-1)G)$th column is in the form of $\boldsymbol{a}(\bar{\theta}_{g_1}, \bar{\theta}_{g_2})$ with $\bar{\theta}_g\triangleq -1+2g/G$, $g_1, g_2=1,2,\ldots,G$, and $\bar{\boldsymbol{x}}^t \in{\mathbb{C}^{G^2}}$ is a sparse vector with $L^t$ nonzero elements corresponding to $\boldsymbol{x}^t$. If $G\gg L^t$, the AoD estimation problem can be transformed into a sparse signal recovery problem, i.e., finding a sparse $\bar{\boldsymbol{x}}^t$ to minimize $\|\boldsymbol{y}^t-\bar{\boldsymbol{A}} \bar{\boldsymbol{x}}^t\|_2$. Classical compressed sensing algorithms, such as the orthogonal matching pursuit (OMP) \cite{lee2016channel}, can be employed to estimate all AoD pairs $\{\theta^t_p, \phi^t_p\}_{p=1}^{L^t}$ corresponding to the columns of $\bar{\boldsymbol{A}}$ with non-zero coefficients in $\bar{\boldsymbol{x}}^t$.

\subsection{AoA Estimation}
Similar to the AoD estimation procedure, in this subsection, the T-MA's position is fixed as $\boldsymbol{t}_M \in\mathcal{C}^t$ and the R-MA is moved over $(N-1)$ different positions denoted by $\{\boldsymbol{r}_n\}_{n=2}^{N}$ with $\boldsymbol{r}_n \triangleq [x^r_n, y^r_n]^T \in\mathcal{C}^r$. For each $\boldsymbol{r}_n$, one pilot is assumed to be sent by the T-MA. The receiver measures the corresponding channels based on the received pilot signals and then estimates the R-MPCs' AoAs. In this step, the received signal vector after $(N-1)$ channel measurements is given by
\begin{align}
	\boldsymbol{y}^r = \sqrt{P}\boldsymbol{F}(\tilde{\boldsymbol{r}})^H \boldsymbol{\Sigma} \boldsymbol{g}(\boldsymbol{t}_M)   + \boldsymbol{z}^r \in \mathbb{C}^{N},
\end{align}
with $\tilde{\boldsymbol{r}} \triangleq \left[\boldsymbol{r}_1, \boldsymbol{r}_2, \ldots, \boldsymbol{r}_N\right] \in \mathbb{R}^{2 \times N}$ and $\boldsymbol{F}(\tilde{\boldsymbol{r}})\triangleq\left[\boldsymbol{f}\left(\boldsymbol{r}_1\right), \boldsymbol{f}\left(\boldsymbol{r}_2\right), \ldots, \boldsymbol{f}\left(\boldsymbol{r}_N\right)\right] \in \mathbb{C}^{L^r \times N}$. $\boldsymbol{z}^r \in \mathbb{C}^{N}$ is the AWGN vector with each entry following $\mathcal{CN}(0,\sigma^2)$.
The receiver-side steering vector is defined as
\begin{align}
	\boldsymbol{b}(\theta^r, \phi^r) \triangleq \left[ e^{-j\frac{2\pi}{\lambda}\tau^r_1(\theta^r, \phi^r)}, e^{-j\frac{2\pi}{\lambda}\tau^r_2(\theta^r, \phi^r)}, \ldots, e^{-j\frac{2\pi}{\lambda}\tau^r_{N}(\theta^r, \phi^r)} \right]^T \in{\mathbb{C}^{N}},
\end{align}
with $\tau^r_n(\theta^r, \phi^r) = x^r_n \theta^r + y^r_n \phi^r$. Then, the receiver-side steering matrix can be written as
\begin{align}
	\boldsymbol{B}(\boldsymbol{\alpha}^r)\triangleq\left[\boldsymbol{b}(\theta^r_1, \phi^r_1), \boldsymbol{b}(\theta^r_2, \phi^r_2), \ldots, \boldsymbol{b}(\theta^r_{L^r}, \phi^r_{L^r})\right] \in \mathbb{C}^{N \times L^r},
\end{align}
with $\boldsymbol{\alpha}^r \triangleq [\theta^r_1, \theta^r_2, \ldots, \theta^r_{L^r}, \phi^r_1, \phi^r_2, \ldots, \phi^r_{L^r}]^T \in{\mathbb{C}^{2L^r}}$. Thus, the received signal vector can be rewritten as
\begin{align}\label{yr2}
	\boldsymbol{y}^r = \sqrt{P}\boldsymbol{B}(\boldsymbol{\alpha}^r) \boldsymbol{\Sigma} \boldsymbol{g}(\boldsymbol{t}_M)   + \boldsymbol{z}^r \triangleq \boldsymbol{B}(\boldsymbol{\alpha}^r) \boldsymbol{x}^r   + \boldsymbol{z}^r,
\end{align}
with $\boldsymbol{x}^r \triangleq \sqrt{P}\boldsymbol{\Sigma} \boldsymbol{g}(\boldsymbol{t}_M) \in{\mathbb{C}^{L^r}}$. Similar to the AoD estimation algorithm in the previous subsection, $\boldsymbol{B}(\boldsymbol{\alpha}^r) \boldsymbol{x}^r$ can be approximated by $\bar{\boldsymbol{B}} \bar{\boldsymbol{x}}^r$, where the $(g_1+(g_2-1)G)$th column of $\bar{\boldsymbol{B}} \in{\mathbb{C}^{N\times{G^2}}}$
is $\boldsymbol{b}(\bar{\theta}_{g_1}, \bar{\theta}_{g_2})$, and $\bar{\boldsymbol{x}}^r \in{\mathbb{C}^{G^2}}$ is a sparse vector with $L^r$ nonzero elements corresponding to $\boldsymbol{x}^r$. If $G\gg L^r$,  all AoA pairs $\{\theta^r_q, \phi^r_q\}_{q=1}^{L^r}$ can be estimated using the OMP algorithm \cite{lee2016channel}, which correspond to the columns of $\bar{\boldsymbol{B}}$ with non-zero coefficients in $\bar{\boldsymbol{x}}^r$.

It is worth noting that the AoD/AoA estimation performance is affected by the T-/R-MA's positions $\{\boldsymbol{t}_m\}_{m=1}^{M}$/$\{\boldsymbol{r}_n\}_{n=1}^{N}$. On one hand, a larger $M$ and $N$ indicate more number of pilots, which can increase the total received signal power for channel reconstruction. On the other hand, the spatial distribution of MA positions impacts the resolution in the angular domain to estimate the AoDs/AoAs. Therefore, we will evaluate the AoD/AoA estimation performance with respect to different MA position setups $\{\{\boldsymbol{t}_m\}_{m=1}^{M}, \{\boldsymbol{r}_n\}_{n=1}^{N}\}$ in Section IV by simulation.

\subsection{PRM Estimation}
Based on the channel measurements in the previous two steps, it is generally not possible to exactly estimate the PRM, $\boldsymbol{\Sigma}$, because the rank of channel measurement matrix is usually smaller than the number of unknown entries in $\boldsymbol{\Sigma}$. Thus, the receiver needs to conduct channel measurements with additional T-/R-MA's positions  $\boldsymbol{t}\in\mathcal{C}^t$ and $\boldsymbol{r}\in\mathcal{C}^r$, so as to estimate $\boldsymbol{\Sigma}$. To reduce the pilot overhead, we exploit the estimated AoDs and AoAs and move the T-MA and R-MA to estimate $\boldsymbol{\Sigma}$ with the minimum number of additional channel measurements or MA's positions. Let $\hat{L}^t$ and $\hat{L}^r$ denote the estimated number of T-MPCs and R-MPCs, respectively. Then, the estimated AoDs and AoAs can be denoted by $\{\hat{\theta}^t_p, \hat{\phi}^t_p\}_{p=1}^{\hat{L}^t}$ and $\{\hat{\theta}^r_q, \hat{\phi}^r_q\}_{q=1}^{\hat{L}^r}$, respectively. Define $\boldsymbol{\gamma}\triangleq \textrm{vec}(\boldsymbol{\Sigma}) \in\mathbb{C}^{L^r L^t}$. Consequently, the received signal vectors in \eqref{yt2} and \eqref{yr2} can be rewritten as
\begin{align}
	\left(\boldsymbol{y}^t\right)^* &\overset{(a)}= \sqrt{P} \left( \boldsymbol{A}(\boldsymbol{\alpha}^t)^* \otimes \boldsymbol{f}(\boldsymbol{r}_1)^H \right) \boldsymbol{\gamma}  + \left(\boldsymbol{z}^t\right)^* \\
	&\triangleq \boldsymbol{\Psi}^t(\boldsymbol{\alpha}^t) \boldsymbol{\gamma}  + \left(\boldsymbol{z}^t\right)^*, \notag\\
	\boldsymbol{y}^r &\overset{(b)}= \sqrt{P} \left( \boldsymbol{g}(\boldsymbol{t}_M)^T \otimes \boldsymbol{B}(\boldsymbol{\alpha}^r) \right)   \boldsymbol{\gamma} + \boldsymbol{z}^r, \notag\\
	&\triangleq \boldsymbol{\Psi}^r(\boldsymbol{\alpha}^r) \boldsymbol{\gamma}  + \boldsymbol{z}^r, \notag
\end{align}
with $\boldsymbol{\Psi}^t(\boldsymbol{\alpha}^t) \triangleq \sqrt{P} \left( \boldsymbol{A}(\boldsymbol{\alpha}^t)^* \otimes \boldsymbol{f}(\boldsymbol{r}_1)^H \right) \in\mathbb{C}^{M \times L^r L^t}$ and $\boldsymbol{\Psi}^r(\boldsymbol{\alpha}^r) \triangleq \sqrt{P} \left( \boldsymbol{g}(\boldsymbol{t}_M)^T \otimes \boldsymbol{B}(\boldsymbol{\alpha}^r) \right) \in\mathbb{C}^{N \times L^r L^t}$. The two equalities marked by $(a)$ and $(b)$ hold since $\textrm{vec}(\boldsymbol{UQV}) = (\boldsymbol{V}^T \otimes \boldsymbol{U})\textrm{vec}(\boldsymbol{Q})$ for $\boldsymbol{U}\in\mathbb{C}^{p \times q}$, $\boldsymbol{Q}\in\mathbb{C}^{q \times r}$, and $\boldsymbol{V}\in\mathbb{C}^{r \times t}$. Denote the number of additional pairs of T-/R-MA positions by $K$, where the $k$th $(k=1,2,\ldots,K)$ positions of the T-MA and R-MA are represented by $\boldsymbol{t}^a_k=[x^{t,a}_k, y^{t,a}_k]^T \in \mathcal{C}^t$ and $\boldsymbol{r}^a_k=[x^{r,a}_k, y^{r,a}_k]^T \in \mathcal{C}^r$, respectively. Then, the received signal for the $k$th additional channel measurement is given by
\begin{align}
	y_k^a = \sqrt{P}\boldsymbol{f}(\boldsymbol{r}^a_k)^H \boldsymbol{\Sigma} \boldsymbol{g}(\boldsymbol{t}^a_k) + z_k^a =\sqrt{P}\left( \boldsymbol{g}(\boldsymbol{t}^a_k)^T \otimes \boldsymbol{f}(\boldsymbol{r}^a_k)^H \right) \boldsymbol{\gamma} + z_k^a \triangleq \psi\left( \boldsymbol{t}^a_k,\boldsymbol{r}^a_k \right)^H \boldsymbol{\gamma} + z_k^a,
\end{align}
where $z_k^a \sim \mathcal{CN}(0,\sigma^2)$ is the AWGN, and $\psi\left( \boldsymbol{t}^a_k,\boldsymbol{r}^a_k \right) \triangleq \sqrt{P}\left( \boldsymbol{g}(\boldsymbol{t}^a_k)^* \otimes \boldsymbol{f}(\boldsymbol{r}^a_k) \right) \in{\mathbb{C}^{L^r L^t}}$. Denote $J\triangleq M+N+K$. The $K$ additional received signals can be stacked with $\boldsymbol{y}^t$ and $\boldsymbol{y}^r$ as
\begin{align}\label{yall}
	\overline{\boldsymbol{y}} = \overline{\boldsymbol{\Psi}}\left( \boldsymbol{\alpha}^t, \boldsymbol{\alpha}^r, \tilde{\boldsymbol{t}}^a, \tilde{\boldsymbol{r}}^a \right)\boldsymbol{\gamma} + \overline{\boldsymbol{z}},
\end{align}
with 
\begin{align}
	\overline{\boldsymbol{y}} &\triangleq [\left(\boldsymbol{y}^t\right)^H, \left(\boldsymbol{y}^r\right)^T, y_1^a, y_2^a, \ldots, y_K^a]^T \in \mathbb{C}^{J}, \\
	\overline{\boldsymbol{z}} &\triangleq [\left(\boldsymbol{z}^t\right)^H, \left( \boldsymbol{z}^r \right)^T, z_1^a, z_2^a, \ldots, z_K^a]^T \in \mathbb{C}^{J}, \notag \\
	\tilde{\boldsymbol{t}}^a &\triangleq \left[\boldsymbol{t}^a_1, \boldsymbol{t}^a_2, \ldots, \boldsymbol{t}^a_K\right] \in \mathbb{R}^{2 \times K}, \notag \\
	\tilde{\boldsymbol{r}}^a &\triangleq \left[\boldsymbol{r}^a_1, \boldsymbol{r}^a_2, \ldots, \boldsymbol{r}^a_K\right] \in \mathbb{R}^{2 \times K}, \notag \\
	\tilde{\boldsymbol{\Psi}}\left( \tilde{\boldsymbol{t}}^a, \tilde{\boldsymbol{r}}^a \right) &\triangleq \left[ \psi\left( \boldsymbol{t}^a_1,\boldsymbol{r}^a_1 \right), \psi\left( \boldsymbol{t}^a_2,\boldsymbol{r}^a_2 \right), \ldots, \psi\left( \boldsymbol{t}^a_K,\boldsymbol{r}^a_K \right) \right] \in{\mathbb{C}^{{L^r L^t}\times K}}, \notag\\
	\overline{\boldsymbol{\Psi}}\left( \boldsymbol{\alpha}^t, \boldsymbol{\alpha}^r, \tilde{\boldsymbol{t}}^a, \tilde{\boldsymbol{r}}^a \right) &\triangleq \left[ \left( \boldsymbol{\Psi}^t(\boldsymbol{\alpha}^t) \right)^T, \left( \boldsymbol{\Psi}^r(\boldsymbol{\alpha}^r) \right)^T, \tilde{\boldsymbol{\Psi}}\left( \tilde{\boldsymbol{t}}^a, \tilde{\boldsymbol{r}}^a \right)^* \right]^T \in{\mathbb{C}^{J\times{L^r L^t}}}. \notag
\end{align}
Define $\hat{\boldsymbol{\alpha}}^t \triangleq [\hat{\theta}^t_1, \hat{\theta}^t_2, \ldots, \hat{\theta}^t_{\hat{L}^t}, \hat{\phi}^t_1, \hat{\phi}^t_2, \ldots, \hat{\phi}^t_{\hat{L}^t}]^T \in{\mathbb{C}^{2\hat{L}^t}}$ and $\hat{\boldsymbol{\alpha}}^r \triangleq [\hat{\theta}^r_1, \hat{\theta}^r_2, \ldots, \hat{\theta}^r_{\hat{L}^r}, \hat{\phi}^r_1, \hat{\phi}^r_2, \ldots, \hat{\phi}^r_{\hat{L}^r}]^T \in{\mathbb{C}^{2\hat{L}^r}}$. Then, $\boldsymbol{\gamma}$ can be estimated by
\begin{align}\label{gamma_hat}
	\hat{\boldsymbol{\gamma}} =\overline{\boldsymbol{\Psi}}\left(\hat{\boldsymbol{\alpha}}^t, \hat{\boldsymbol{\alpha}}^r, \tilde{\boldsymbol{t}}^a, \tilde{\boldsymbol{r}}^a\right)^\dagger \overline{\boldsymbol{y}}.
\end{align}
Since $\textrm{rank}\left(\overline{\boldsymbol{\Psi}}\left(\hat{\boldsymbol{\alpha}}^t, \hat{\boldsymbol{\alpha}}^r, \tilde{\boldsymbol{t}}^a, \tilde{\boldsymbol{r}}^a\right)\right) \leq \hat{L}_r + \hat{L}_t + K$, it requires $K\geq \hat{L}^r \hat{L}^t - \hat{L}_r - \hat{L}_t$ such that the measurement matrix $\overline{\boldsymbol{\Psi}}\left(\hat{\boldsymbol{\alpha}}^t, \hat{\boldsymbol{\alpha}}^r, \tilde{\boldsymbol{t}}^a, \tilde{\boldsymbol{r}}^a\right)$ has a full-column rank.

The additional T-MA's and R-MA's positions affect the estimation performance of $\boldsymbol{\gamma}$. If $\overline{\boldsymbol{\Psi}}\left(\hat{\boldsymbol{\alpha}}^t, \hat{\boldsymbol{\alpha}}^r, \tilde{\boldsymbol{t}}^a, \tilde{\boldsymbol{r}}^a\right)$ is not well-conditioned, it will amplify the noise and reduce the estimation accuracy. Therefore, we optimize $\tilde{\boldsymbol{t}}^a$ and $\tilde{\boldsymbol{r}}^a$ to minimize the condition number of $\overline{\boldsymbol{\Psi}}\left(\hat{\boldsymbol{\alpha}}^t, \hat{\boldsymbol{\alpha}}^r, \tilde{\boldsymbol{t}}^a, \tilde{\boldsymbol{r}}^a\right)$, which is formulated as
\begin{align}\label{condition}
	\underset{\tilde{\boldsymbol{t}}^a, \tilde{\boldsymbol{r}}^a} {\min} \quad & \kappa\left( \overline{\boldsymbol{\Psi}}\left(\hat{\boldsymbol{\alpha}}^t, \hat{\boldsymbol{\alpha}}^r, \tilde{\boldsymbol{t}}^a, \tilde{\boldsymbol{r}}^a\right) \right) \\
	\text{s.t.} \quad & \tilde{\boldsymbol{t}}^a \in \mathcal{C}^t, \notag\\
	& \tilde{\boldsymbol{r}}^a \in \mathcal{C}^r. \notag
\end{align}
Problem \eqref{condition} is difficult to solve because the objective function is highly non-convex with respect to $\tilde{\boldsymbol{t}}^a$ and $\tilde{\boldsymbol{r}}^a$. Since $\tilde{\boldsymbol{t}}^a$ and $\tilde{\boldsymbol{r}}^a$ are real vectors and the constraints of problem \eqref{condition} are linear, the local optimum of problem \eqref{condition} can be efficiently found based on the off-the-shelf fmincon function in MATLAB. Finally, the estimated $h(\boldsymbol{t}, \boldsymbol{r})$ can be expressed as 
\begin{equation}\label{Hhat}
	\hat{h}(\boldsymbol{t}, \boldsymbol{r}) = \hat{\boldsymbol{f}}(\boldsymbol{r})^H \hat{\boldsymbol{\Sigma}} \hat{\boldsymbol{g}}(\boldsymbol{t}),
\end{equation}
where $\hat{\boldsymbol{\Sigma}}$ is obtained by devectorizing $\hat{\boldsymbol{\gamma}}$ to an $\hat{L}_r\times \hat{L}_t$ matrix, $\hat{\boldsymbol{g}}(\boldsymbol{t}) \in{\mathbb{C}^{\hat{L}_t}}$ and $\hat{\boldsymbol{f}}(\boldsymbol{r}) \in{\mathbb{C}^{\hat{L}_r}}$ are the estimated T-FRV and R-FRV, with $\hat{\boldsymbol{g}}(\boldsymbol{t})[p]=e^{j\frac{2\pi}{\lambda}(x^t \hat{\theta}^t_p + y^t \hat{\phi}^t_p)}$ and $\hat{\boldsymbol{f}}(\boldsymbol{r})[q]=e^{j\frac{2\pi}{\lambda}(x^r \hat{\theta}^r_q + y^r \hat{\phi}^r_q)}$, respectively.

\begin{algorithm}[!t]
	\begin{algorithmic}[1]
		\STATE \emph{Input:} $\boldsymbol{y}^t$, $\boldsymbol{y}^r$, $K$, $\hat{L}_r$, $\hat{L}_t$.
		\STATE Obtain the estimated AoDs $\{\hat{\theta}^t_p, \hat{\phi}^t_p\}_{p=1}^{\hat{L}^t}$ and estimated AoAs $\{\hat{\theta}^r_q, \hat{\phi}^r_q\}_{q=1}^{\hat{L}^r}$ by using the OMP method at the receiver.
		\STATE Obtain $\tilde{\boldsymbol{t}}^a$ and $\tilde{\boldsymbol{r}}^a$ at the receiver by solving \eqref{condition}.
		\STATE The receiver feeds back $\tilde{\boldsymbol{t}}^a$ to the transmitter for changing T-MA positions, and then obtains $\overline{\boldsymbol{y}}$ via \eqref{yall}.
		\STATE Compute $\hat{\boldsymbol{\gamma}}$ via \eqref{gamma_hat} at the receiver.
		\STATE Obtain $\hat{h}(\boldsymbol{t}, \boldsymbol{r})$ via \eqref{Hhat} at the receiver.
		
		\STATE \emph{Output:} $\hat{h}(\boldsymbol{t}, \boldsymbol{r})$.
	\end{algorithmic}
	\caption{STRCS-based Channel Reconstruction Algorithm}
	\label{alg1}
\end{algorithm}

The detailed steps of the proposed STRCS-based channel reconstruction  algorithm are summarized in Algorithm~\ref{alg1}. Specifically, at step 2, we use the OMP method to obtain the estimated AoDs $\{\hat{\theta}^t_p, \hat{\phi}^t_p\}_{p=1}^{\hat{L}^t}$ and estimated AoAs $\{\hat{\theta}^r_q, \hat{\phi}^r_q\}_{q=1}^{\hat{L}^r}$ at the receiver based on the methods given in Sections III-A and III-B, respectively. Then, at step 3, we obtain $\overline{\boldsymbol{\Psi}}\left(\hat{\boldsymbol{\alpha}}^t, \hat{\boldsymbol{\alpha}}^r, \tilde{\boldsymbol{t}}^a, \tilde{\boldsymbol{r}}^a\right)$ based on $\{\hat{\theta}^t_p, \hat{\phi}^t_p\}_{p=1}^{\hat{L}^t}$ and $\{\hat{\theta}^r_q, \hat{\phi}^r_q\}_{q=1}^{\hat{L}^r}$, and then solve problem \eqref{condition} to obtain $\tilde{\boldsymbol{t}}^a$ and $\tilde{\boldsymbol{r}}^a$. Finally, $\hat{\boldsymbol{\gamma}}$ is computed by LS via \eqref{gamma_hat} at the receiver, and then $\hat{h}(\boldsymbol{t}, \boldsymbol{r})$ is reconstructed based on the estimated FRI via \eqref{Hhat}.

Next, we analyze the computational complexity of the proposed STRCS-based algorithm. The complexity of the AoD and AoA estimation using OMP is $\mathcal{O}(M \hat{L}^t G^2)$ and $\mathcal{O}(N \hat{L}^r G^2)$, respectively \cite{lee2016channel}. The complexity of solving problem \eqref{condition} is $\mathcal{O}(K^{3.5}\log(1/\epsilon))$ with accuracy $\epsilon$ for the interior-point method \cite{fu2021reconf}. The complexity to obtain $\hat{\boldsymbol{\gamma}}$ by LS is $\mathcal{O}((\hat{L}^r \hat{L}^t)^2 J)$. Thus, the total
complexity of Algorithm~\ref{alg1} is $\mathcal{O}(M \hat{L}^t G^2 + N \hat{L}^r G^2 + K^{3.5}\log(1/\epsilon) + (\hat{L}^r \hat{L}^t)^2 J)$, which is polynomial over $M$, $N$, $\hat{L}^t$, $\hat{L}^r$, $G$, $K$, and $J$.

\section{Numerical Results}
This section presents simulation results for validating our proposed STRCS-based channel reconstruction method for MA systems. In the simulation, $\mathcal{C}^t$ and $\mathcal{C}^r$ are both set as square areas with $A=4\lambda$. The numbers of T-MPCs and R-MPCs are set as $L^t=L^r=3$ \cite{3gpp2016study}. As such, PRM is a square matrix with $\boldsymbol{\Sigma}[p,p]\sim \mathcal{CN}(0, \eta/((\eta+1)L^r))$ and $\boldsymbol{\Sigma}[p,q]\sim \mathcal{CN}(0, 1/((\eta+1)(L^r-1)L^r))$ for $\{(p,q) | p=1,2,\ldots,L^r; q=1,2,\ldots,L^r; p\neq q\}$, where $\eta$ is set as $1$ denoting the ratio of the average power for diagonal elements to that for non-diagonal elements \cite{zhu2022modeling}. The elevation and azimuth AoDs/AoAs are random variables following uniform distribution over $[0, \pi]$. Due to the normalized channel power, the average receive SNR is defined as $P/\sigma^2$. We set $N=M=256$ as the number of T-/R-MA positions in the first two steps of the proposed STRCS method. The OMP method is used to estimate AoDs/AoAs, with $\hat{L}^r = L^r+1$, $\hat{L}^t = L^t+1$, and $G=200$ \cite{wang2020compressed}. We present simulation results based on averaging over $10^4$ independent channel realizations. To investigate the channel reconstruction performance, $\mathcal{C}^t$ and $\mathcal{C}^r$ are uniformly divided into multiple grids, with the distance of adjacent grids being $\Delta\triangleq\lambda/5$. Then, there are total $D\triangleq(A/\Delta)^2$ grids in $\mathcal{C}^t$/$\mathcal{C}^r$, with their center locations denoted by $\{\boldsymbol{t}_u^s\}_{u=1}^{D}$/$\{\boldsymbol{r}_v^s\}_{v=1}^{D}$. Then, Let $\boldsymbol{H} \in{\mathbb{C}^{D\times{D}}}$ and $\hat{\boldsymbol{H}} \in{\mathbb{C}^{D\times{D}}}$ denote the channel matrix from all the centers of transmit grids to all the centers of receive grids and the reconstruction of $\boldsymbol{H}$, i.e., $\boldsymbol{H}[v,u]=h(\boldsymbol{t}_u^s, \boldsymbol{r}_v^s)$ and $\hat{\boldsymbol{H}}[v,u]=\hat{h}(\boldsymbol{t}_u^s, \boldsymbol{r}_v^s)$, respectively. Then, the normalized mean square error (NMSE) for channel reconstruction is defined by
\begin{align}
	\textrm{NMSE} = \mathbb{E}\left\{ \frac{\|\boldsymbol{H}-\hat{\boldsymbol{H}}\|_F^2}{\|\boldsymbol{H}\|_F^2} \right\},
\end{align}
which is averaged over different channel realizations.

\begin{figure}
	\centering
	\includegraphics[width=100mm]{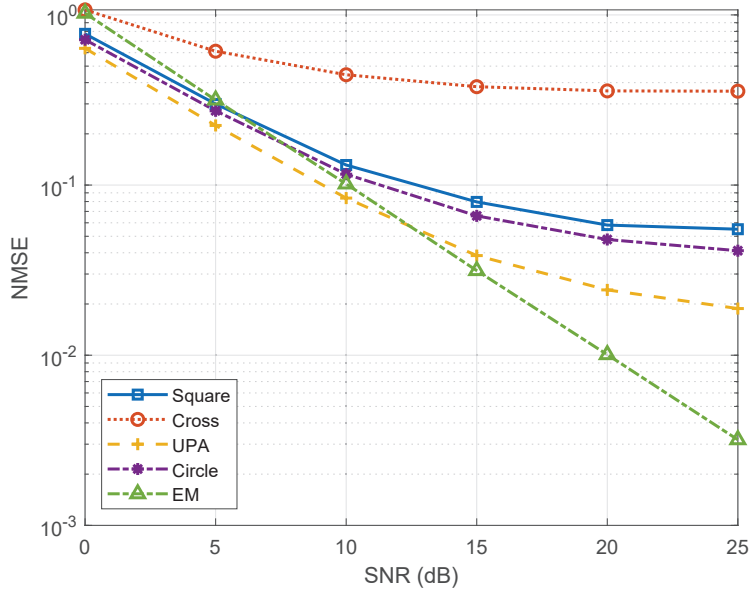}
	\caption{NMSE versus SNR for different MA position setups in the first two steps.}
	\label{SNR}
\end{figure}

In Fig.~\ref{SNR}, we show the NMSE versus average receive SNR. We use $K=16$ additional pairs of T-/R-MA positions in the third step, which are selected by solving \eqref{condition}. In particular, we consider four different MA position setups in the first two steps: 1) \textbf{Square-shape}: $\{\boldsymbol{t}_m\}_{m=1}^{M}$/$\{\boldsymbol{r}_n\}_{n=1}^{N}$ forms four lines along the contour of $\mathcal{C}^t$/$\mathcal{C}^r$, with spacing $4A/M$ between adjacent positions; 2) \textbf{Cross-shape}: $\{\boldsymbol{t}_m\}_{m=1}^{M}$/$\{\boldsymbol{r}_n\}_{n=1}^{N}$ forms two perpendicular lines intersecting at the center of $\mathcal{C}^t$/$\mathcal{C}^r$, with spacing $2A/(M-1)$ between adjacent positions; 3) 
\textbf{Uniform planar array (UPA)-shape}: $\{\boldsymbol{t}_m\}_{m=1}^{M}$/$\{\boldsymbol{r}_n\}_{n=1}^{N}$ forms the UPA shape, with spacing $A/(\sqrt{M}-1)$ between adjacent positions; and 4) \textbf{Circle-shape}: $\{\boldsymbol{t}_m\}_{m=1}^{M}$/$\{\boldsymbol{r}_n\}_{n=1}^{N}$ forms an inscribed circle in $\mathcal{C}^t$/$\mathcal{C}^r$, with spacing $2A\sin(2\pi/M)$ between adjacent positions. Besides, we consider the \textbf{exhaustive measurement (EM)} benchmark by directly measuring all $D\times D$ channels of $\boldsymbol{H}$. As shown in Fig.~\ref{SNR}, the proposed channel reconstruction method can be applied to different MA position setups. Besides, the UPA-shaped position setup can achieve lower NMSE as compared to the other setups. This is because the UPA geometry can extract more spatial information to distinguish with different MPCs' AoDs/AoAs. For $\textrm{SNR}=25$ dB, the UPA-shaped position setup achieves $65.8\%$, $94.7\%$, and $54.3\%$ performance improvements over the square-shaped, cross-shaped, and circle-shaped position setups, respectively. Moreover, the proposed method with UPA-shaped position setup has lower NMSE than the EM method when $\textrm{SNR}<12$ dB, while its number of total measurements is $M+N-1+K=527$, which is much smaller than that of the EM method, i.e., $D^2=1.6\times 10^5$.

\begin{figure}
	\centering
	\includegraphics[width=100mm]{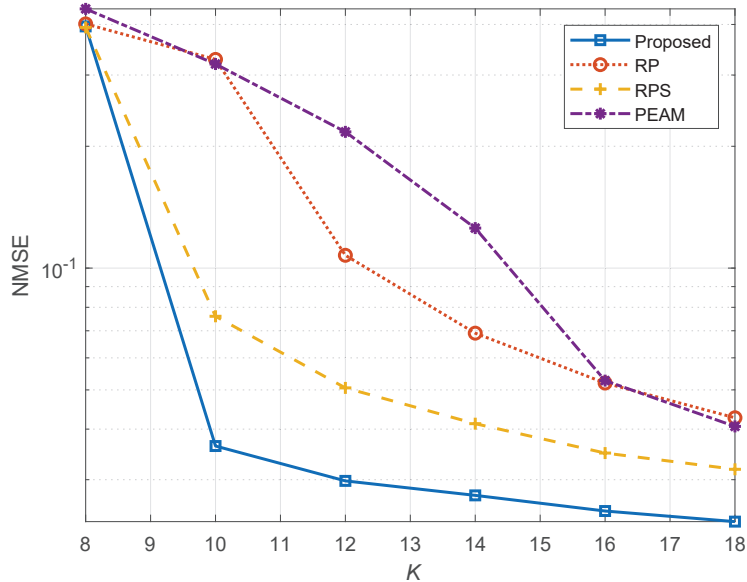}
	\caption{NMSE versus $K$ for different MA position selection methods in the third step.}
	\label{K}
\end{figure}

In Fig.~\ref{K}, we verify the performance of channel reconstruction under different  MA position selection methods in the third step of the proposed STRCS method. We set $\textrm{SNR}=20$ dB and use the same UPA-shaped MA position setup for AoD/AoA estimation in the first two steps. The following benchmark schemes are considered for the third step: 1) \textbf{Random position (RP)}: Randomly generate $\tilde{\boldsymbol{t}}^a\in \mathcal{C}^t$ and $\tilde{\boldsymbol{r}}^a\in \mathcal{C}^r$; 2) \textbf{Random position selection (RPS)}: Randomly generate $\tilde{\boldsymbol{t}}^a\in \mathcal{C}^t$ and $\tilde{\boldsymbol{r}}^a\in \mathcal{C}^r$
for $100$ independent realizations, and select $\tilde{\boldsymbol{t}}^a$ and $\tilde{\boldsymbol{r}}^a$ with the lowest $\kappa\left( \overline{\boldsymbol{\Psi}}\left(\hat{\boldsymbol{\alpha}}^t, \hat{\boldsymbol{\alpha}}^r, \tilde{\boldsymbol{t}}^a, \tilde{\boldsymbol{r}}^a\right) \right)$; 3) \textbf{PRM estimation only by additional measurements (PEAM)}: Obtain $\tilde{\boldsymbol{t}}^a$ and $\tilde{\boldsymbol{r}}^a$ by minimizing $\kappa\left(\tilde{\boldsymbol{\Psi}}\left( \tilde{\boldsymbol{t}}^a, \tilde{\boldsymbol{r}}^a \right)\right)$, and then estimate $\boldsymbol{\gamma}$ by $\hat{\boldsymbol{\gamma}} =\tilde{\boldsymbol{\Psi}}\left( \tilde{\boldsymbol{t}}^a, \tilde{\boldsymbol{r}}^a \right)^\dagger [y_1^a, y_2^a, \ldots, y_K^a]^T$ without using the channel measurements in the first two steps. It is observed that with the same number of additional channel measurements in the third step, i.e., $K$, the proposed scheme by solving \eqref{condition} can achieve much smaller NMSE as compared to other benchmark schemes. For the case with $K=10$, the proposed scheme has $88.9\%$, $52.2\%$, and $88.6\%$ performance improvements over the RP, RPS, and PEAM schemes, respectively. Furthermore, it is shown in Fig.~\ref{K} that the proposed scheme outperforms the RP and RPS schemes, indicating that optimizing the T-/R-MA's positions can yield a more well-conditioned measurement matrix such that higher estimation accuracy of the PRM can be achieved.

\section{Conclusions}
In this letter, we proposed a new channel reconstruction   method for MA systems by exploiting the sparse representation of the channel responses in the given transmitter/receiver region in terms of MPCs. To reduce the pilot overhead and computational complexity for channel estimation, the FRI in the angular domain,
including the AoDs/AoAs and complex coefficients of all significant MPCs were sequentially estimated based on finite channel measurements taken at random/selected locations by the MA in the given region. Numerical results demonstrated that the proposed method achieves high channel reconstruction  accuracy with only moderate pilot overhead required, which is suitable for wireless applications with slow-varying channels.

\bibliographystyle{IEEEtran}
\bibliography{IEEEabrv,IEEEexample}
\end{document}